\documentclass[prl, twocolumn, nofootinbib, showpacs, groupedaddress]{revtex4} 
\usepackage{graphicx}  
\usepackage{dcolumn}   
\usepackage{bm}        
\usepackage{amssymb}   
\usepackage{epsfig}
\usepackage{amsmath} 

\begin{document}
\title{Testing the Self-Consistency of the Excursion Set Approach}

\author{I. Achitouv\footnote{achitouv@usm.uni-muenchen.de}$^{1,2,3}$, Y. Rasera$^{3}$, R.K. Sheth$^{4,5}$, P.S. Corasaniti$^{3}$}
\affiliation{$^1$Ludwig-Maximilians-Universitat M\"unchen, Universit\"ats-Sternwarte M\"unchen, Scheinerstr. 1, D-81679 M\"unchen, Germany\\
$^2$Excellence Cluster Universe, Boltzmannstr. 2, D-85748 Garching bei M\"unchen, Germany\\
$^3$Laboratoire Univers et Th\'eories (LUTh), UMR 8102 CNRS, Observatoire de Paris, Universit\'e Paris Diderot, 5 Place Jules Janssen, 92190 Meudon, France\\
$^4$The Abdus Salam International Center for Theoretical Physics, Strada Costiera, 11, Trieste 34151, Italy\\
$^5$Center for Particle Cosmology, University of Pennsylvania, 209 S. 33rd St., Philadelphia, PA 19104, USA
}

\date{\today}

\begin{abstract} 
The excursion set approach provides a framework for predicting how the abundance of dark matter halos depends on the initial conditions.  A key ingredient of this formalism is the specification of a critical overdensity threshold (barrier) which protohalos must exceed if they are to form virialized halos at a later time. However, to make its predictions, the excursion set approach explicitly averages over all positions in the initial field, rather than the special ones around which halos form, so it is not clear that the barrier has physical motivation or meaning. In this letter we show that once the statistical assumptions which underlie the excursion set approach are considered a drifting diffusing barrier model does provide a good self-consistent description both of halo abundance as well as of the initial overdensities of the protohalo patches.

\end{abstract}

\maketitle
Building upon the seminal work of Press \& Schechter \cite{PS1974}, the excursion set approach \cite{BCEK1991,LC1993,MW1996,RKS1998} is the most widely developed formalism for estimating how halo abundances depend on the background cosmology.  Rather than directly predicting the comoving density of halos $dn/dM$ in the mass range $[M,M+dM]$, it provides an estimate of the mass fraction in halos, $(M/\rho)\, dn/dM$, where $\rho$ is the comoving background density.

The approach uses the fact that, at any randomly chosen point in the initial density fluctuation field, the overdensity $\delta$ performs a random walk as a function of the smoothing scale $R$.  Since initial fluctuations are small, this scale can be associated with a mass, $M = \rho V(R)$, where $V(R)\propto R^3$ is the comoving volume of the filter which was used to smooth the field.  The approach assumes that the mass fraction $(M/\rho)\, dn/dM$ equals the fraction of random walk trajectories ${\cal F}[M(R)]$ for which $R$ is the largest scale on which $\delta(R)$ exceeds a critical threshold value, $\delta_{th}$.  

Thus, the approach has two key ingredients.  The first is $\delta_{th}$, which is assumed to be simply related to the physics of halo formation.  E.g., the simplest excursion set formulae equate $\delta_{th}$ with the value $\delta_{sc}$ associated with the collapse of an initially homogeneous overdense sphere \cite{GG1972}. The second is the ensemble of walks which must cross $\delta_{th}$.  

The full set of walks, associated with all positions in the initial field, is determined by the statistics of the initial density fluctuation field and the filter function $W$. How one should average over this ensemble is not usually stated explicitly.  The excursion set approach implicitly assumes that the appropriate average is over the entire ensemble.  However, halos have been shown to collapse around special positions in the initial density field \cite{SMT2001}; e.g., protohalos are often local peaks in the smoothed field $\delta(R)$ \cite{LP2011}.  Hence, if one is not averaging over the special subset of walks around which collapse occurs, and for which the physics is presumably the simplest, then it is not \textit{a priori} guaranteed that $\delta_{th}$ should be simply related to $\delta_{sc}$. 
This and other issues have led to the question of whether or not the excursion set approach is self-consistent \cite{SDMW1996, SMT2001, RKTZ2009}.  

In this Letter we show that if one considers the statistics of all (rather than special) walks having to cross an effective boundary -- one which may not be the same as that associated with models of halo collapse -- then one can indeed build a self-consistent excursion set theory.

We perform our analysis using halos identified in N-body simulations of the DEUS consortium \cite{DEUS}. These are described in \cite{Rasera,Ali,Courtin} and are publicly available through the DEUVO database\footnote{http://www.deus-consortium.org/deuvo/}. The simulations, of a $\Lambda$CDM model calibrated to the WMAP-5yr data with $\sigma_8=0.79$, have box-lengths of 162, 648 and 2592$h^{-1}$Mpc respectively with $1024^3$ particles.  They were realized using the RAMSES code \cite{Teyssier2002}; halos were found using an FoF finder with link-length $b=0.2$.

\medskip



Most excursion set analyses simplify considerably if one works, not with the smoothing scale $R$ or the mass $M$, but with the variance $S$ of the walk height:
 $\langle\delta^2\rangle\equiv\sigma^2\equiv S$ where 
 $S(R)=\frac{1}{2\pi^2}\int dk\,k^2P(k)\tilde{W}^2(k,R)$. 
For $\Lambda$CDM, $S$ is a monotonically decreasing function of $R$ or $M$.  
Since $S$ is a deterministic function of $M$, if one thinks of the walk height $\delta$ as being the sum of many steps, then the problem is to find the smallest $S$ at which 
\begin{equation}\label{deterministic}
 \delta\ge\delta_{th}, 
\end{equation}
where we will sometimes call $\delta_{th}$ the `barrier height'.

In the simplest version of the approach, $\delta_{th}$ is assumed to be a constant, independent of spatial position and smoothing scale \cite{BCEK1991}.  The next level of complication allows $\delta_{th}$ to be a function of scale $S$, but not of position \cite{SMT2001}.  Finally, one may imagine that $\delta_{th}$ depends both on $S$ and on position \cite{SMT2001,CL2001,ST2002}.  Whereas the simplest version is associated with the spherical collapse model, the latter two arise naturally in ellipsoidal collapse models where the initial overdensity is not the only parameter which determines the collapse \cite{D1970,M1995,BM1996,ATA1997,LS1998,SMT2001}.  In such models, one may write 
\begin{equation}\label{eq2}
 \delta\ge\delta_{th}\equiv\bar{B}(S) + B, 
\end{equation}
where $\bar{B}(S)$ is a deterministic function of $S$ alone which encapsulates the main features of the ellipsoidal collapse, and $B$ is stochastic. 
Though the original ellipsoidal collapse model suggests that $B$ is non-Gaussian \cite{ST2002,D2008,SCS2012}, in what follows, we will assume that B is independent of $\delta$ and has a Gaussian PDF, with zero mean and variance $D_B S$. In fact, such a simple double-Gaussian drifting diffusing barrier model posses a number of interesting properties \cite{MR2010b,CA2011a,CA2011b,CS2013}.

Written this way, one may think of the right-hand-side of Eq.~(\ref{eq2}) as being a stochastic barrier which $\delta$ must cross.  
The associated first crossing distribution ${\cal F}(M)dM = {\cal F}(S)dS$ is trivially related to the one in which there is no stochasticity by noting that like $\delta$, $\delta-B$ is also Gaussian with mean zero:  only the variance is $S\to S(1+D_B)$.  However, the first crossing distribution of a drifting deterministic barrier is, in general, a rather complicated function of $\bar{B}(S)$ and $S$ which depends on the smoothing filter $W$.  

The dependence on $W$ is easy to appreciate. For Gaussian initial conditions, and a filter which allows one k-mode at a time (hereafter sharp-k), the steps in a walk are uncorrelated with one another, whereas for a filter which is a top-hat in real space (hereafter sharp-x) the steps are correlated.  The sharp-x filter is the one most often used to define halos and to model the physics of halo formation and collapse, but the associated correlations between steps complicate the excursion set estimate of ${\cal F}(S)$. These can be accounted-for using numerical (Monte Carlo) \cite{BCEK1991} or perturbative computational methods \cite{MR2010a,CA2011a}, or other approaches \cite{PH1990,PLS2012,MS2012,MS2013}.
\medskip

There is a sense in which stochastic barrier models such as this one differ fundamentally from a deterministic barrier model. Namely, the condition $\delta - B = \bar{B}(S)$ may be satisfied at many different values of $\delta$ \cite{ST2002}. We will use $\delta_{1x}$ to denote the value of $\delta$ at first crossing and $\Pi(\delta_{1x},S)$ its distribution.

If $B$ is deterministic (i.e. $D_B=0$), then $\Pi(\delta_{1x},S)$ is a Dirac-delta function centered on $\bar{B}$.  
We can derive an expression for $\Pi(\delta_{1x},S)$ by first noticing that in the coordinate system $(g_1,g_2)$ where $g_2=B/\sqrt{D_B}$ and $g_1=\delta$, the crossing condition $\delta = \bar{B}(S) + B$ defines a line.  It is in this sense that it is better to not think of the `barrier' to be crossed as being `stochastic', but as a $2$-dimensional random walk crossing a deterministic barrier (the line). Moreover the fact that this barrier is a line means that it is {\em much} more convenient to think of the $2$-D walk as taking steps which are parallel and perpendicular to the barrier. Therefore, one should rotate the coordinate system to:  
\begin{equation}\label{sysEq1}
\begin{split}
 &g_+ = (g_1-\sqrt{D_B}g_2)/\sqrt{1+D_B}\\
 &g_- = (g_1\sqrt{D_B}+g_2)/\sqrt{1+D_B} .
\end{split}
\end{equation}
Notice that $\langle g_+\rangle = \langle g_{-}\rangle = 0$ and  
$\langle g_+^2\rangle = \langle g_{-}^2\rangle = S$ and $\langle g_+ g_-\rangle = 0$. At $\delta_{1x} = \bar{B}(S) + B = \bar{B}(S) + g_2\sqrt{D_B}$ 
\begin{equation}\label{sysEq2}
\begin{split}
 &g_+ = \bar{B}(S)/\sqrt{1+D_B}\\
 &g_- = \bar{B}(S)\sqrt{D_B/(1+D_B)} + g_2\sqrt{1+D_B},
\end{split}
\end{equation}
making 
 $g_2 = g_-/\sqrt{1+D_B} - \bar{B}(S)\sqrt{D_B/(1+D_B)}$
and hence 
\begin{equation}
 \delta_{1x} =  \bar{B}(S) + g_2\sqrt{D_B} 
            = \frac{ \bar{B}(S)}{1+D_B} + g_-\sqrt{\frac{D_B}{1+D_B}}.
\end{equation}
If we define $\mu_{1x}\equiv \bar{B}(S)/(1+D_B)$, then this shows that the distribution of $\delta_{1x} - \mu_{1x}$ is just a rescaled version of that of $g_-$ subject to the constraint that $g_+$ crossed $\bar{B}(S)/\sqrt{1+D_B}$ for the first time on scale $S$.  

When both $\delta$ and $B$ have been smoothed with the same filter then $g_-$ is a zero-mean Gaussian with variance $S$ \cite{CS2013}, so
\begin{equation}
 \Pi(\delta_{1x},S) = \dfrac{e^{-\frac{(\delta_{1x}-\mu_{1x})^2}{2 D_B^{\rm{eff}}S}}}{\sqrt{2 \pi D_B^{\rm{eff}}S}} \quad {\rm where}\quad D_B^{\rm{eff}} = \frac{D_B}{1+D_B}.
 \label{pd1x}
\end{equation}
But if the two filters are different, then one may expect a small correction to the above formula.  Note that the argument above is independent of the specific choice of $\bar{B}(S)$, so it is valid for any $\bar{B}(S)$, not just the special case where the barrier is a constant independent of $S$.  In this respect, $\Pi(\delta_{1x},S)$ is much simpler than is ${\cal F}(S)$ itself.  

Notice that $\bar{B}(S)$ and $D_B$ are the same quantities which appear in the first crossing distribution.  Therefore, if the excursion set approach is self-consistent, then $\Pi(\delta_{1x},S)$, with parameters calibrated from fitting $f(S)$ to halo counts in N-body simulations, should provide a good description of the distribution of $\delta_{1x}$ measured in the same simulations. 

\medskip

Previous work \cite{CA2011a,CA2011b,AC2012a} has shown that ${\cal F}(S)$, associated with a linear barrier 
\begin{equation}
 \label{linear}
 \bar{B}(S) = \delta_{sc} + \beta S, 
\end{equation}
provides a good description of the mass fraction in halos when $\delta$ has sharp-x smoothing and $B$ is sharp-k smoothed, so we will use this two-filter set-up in what follows. This assumes that the collapse conditions at different scales are uncorrelated. Note that this means the smoothing scales associated with the filters $R_k$ and $R_{x}$ have been matched by requiring $\langle B^2(R_k)\rangle = D_B\,\langle\delta^2(R_{x})\rangle$. This makes the scale of the sharp-k filter (which we use for $B$) about a factor of 2 smaller than that for the sharp-x filter (which we use for $\delta$).

The multiplicity function $f(\sigma)\equiv 2 S\mathcal{F}(S)$ obtained using the perturbative path-integral method of \cite{MR2010a} is given by

\begin{equation}
f(\sigma) = f_0(\sigma) + f_{1,\beta=0}^{m-m}(\sigma)
          + f_{1,\beta^{(1)}}^{m-m}(\sigma) + f_{1,\beta^{(2)}}^{m-m}(\sigma)\label{ftot}
\end{equation}
with
\begin{equation*}
 f_0(\sigma) = \frac{\delta_{sc}}{\sigma}\sqrt{\frac{2 a}{\pi}}\, e^{-a\bar{B}^2/2\sigma^2},
\end{equation*}
\begin{equation*}
f_{1,\beta=0}^{m-m}(\sigma)=-\tilde{\kappa}\dfrac{\delta_{sc}}{\sigma}\sqrt{\frac{2a}{\pi}}\left[e^{-\frac{a \delta_{sc}^2}{2\sigma^2}}-\frac{1}{2} \Gamma\left(0,\frac{a\delta_{sc}^2}{2\sigma^2}\right)\right],
\end{equation*}
\begin{equation*}
f_{1,\beta^{(1)}}^{m-m}(\sigma)=- a\,\delta_{sc}\,\beta\left[\tilde{\kappa}\,\text{Erfc}\left( \delta_{sc}\sqrt{\frac{a}{2\sigma^2}}\right)+ f_{1,\beta=0}^{m-m}(\sigma)\right],
\end{equation*}
\begin{equation*}\label{beta2}
f_{1,\beta^{(2)}}^{m-m}(\sigma)=-a\,\beta\left[\frac{\beta}{2} \sigma^2 f_{1,\beta=0}^{m-m}(\sigma)+\delta_{sc} \,f_{1,\beta^{(1)}}^{m-m}(\sigma)\right],
\end{equation*}
where $\tilde{\kappa}=a\,\kappa = \kappa/(1+D_B)$ \cite{CA2011b}\footnote{Our expression for $f_{1,\beta^{(2)}}^{m-m}$ has a term $\mathcal{O}(\beta^2)$ corrected in \cite{AC2012b}.}. Note that in the case of a deterministic linear barrier ($D_B=0$) and a sharp-k filter ($\kappa=0$) we recover the solution of \cite{RKS1998}.

\begin{figure}
\begin{center}
\includegraphics[scale=0.35]{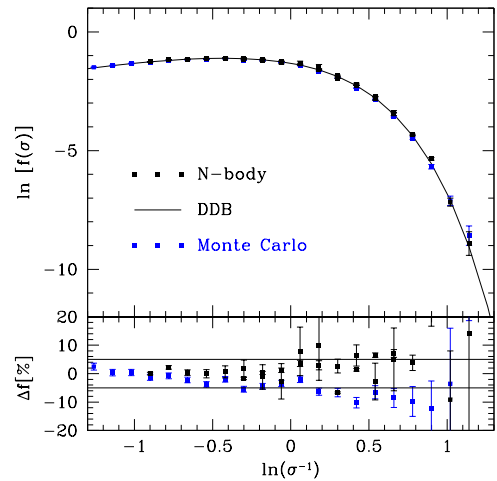}
\caption{(Upper panel) Abundance of FoF ($b=0.2$) halos from N-body simulations (black symbols), and $f(\sigma)$ of Eq.~(\ref{ftot}) (black solid line) with $\beta=0.12$ and $D_B=0.40$.  Blue symbols show the corresponding excursion set first crossing distribution obtained by Monte-Carlo simulations of the random walks. (Lower panel) Relative difference with respect to Eq.~(\ref{ftot}). Thin black lines indicate $5\%$ deviations.}\label{fig1}
\end{center}
\end{figure}

For $\Lambda$CDM, $\delta_c=1.673$ and $\kappa=0.465$, so Eq.~(\ref{ftot}) has two free parameters:  $\beta$ and $D_B$.  We determine these by fitting Eq.~(\ref{ftot}) to the DEUS halo counts, finding $\beta=0.12\pm 0.1$ and $D_B=0.40\pm0.03$.
Fig.~\ref{fig1} shows that, for this pair of values, the discrepancy between Eq.~(\ref{ftot}) and the numerical data at $z=0$ is $\sim 5\%$, which is consistent with \cite{CA2011a,CA2011b,AC2012a} and with numerical uncertainties on the mass function \cite{Courtin}. Fig.~\ref{fig1} also shows that Eq.~(\ref{ftot}) provides a similarly good description of the associated first crossing distribution obtained by direct Monte-Carlo simulation of the sharp-x $\delta$ and sharp-k $B$ walks consistently with the findings of \cite{CA2011b}. 
Note in particular that $1/\sqrt{1+D_B} = \sqrt{0.71}$ explains the appearance of the numerical factor multiplying $\delta_{sc}$ in \cite{ST1999}.

\medskip

We now turn to the question:  Is this a self-consistent description? If it is, then $\Pi(\delta_{1x},S)$ measured for the same walks used to make in Fig.~\ref{fig1} should provide a good description of the distribution of $\delta_{1x}$ measured in the DEUS simulations. To estimate this in the simulations we select a random particle for each halo of mass $M(R)$ and evaluate the overdensity $\delta_{1x}$ within a sphere of radius $R=(3 M/4\pi\rho)^{1/3}$ around it in the initial conditions. Choosing a random halo particle (rather than the one at the protohalo center, say) is crucial, since this makes the measurement to correspond to the quantity which the usual excursion set calculation returns \cite{SDMW1996}:  i.e., an average over all positions in the initial field. In contrast, the particle which lies closest to the initial center of mass represents a special subset of all walks \cite{SMT2001}. We will consider such walks shortly.

\begin{figure}
\begin{center}
\includegraphics[scale=0.35]{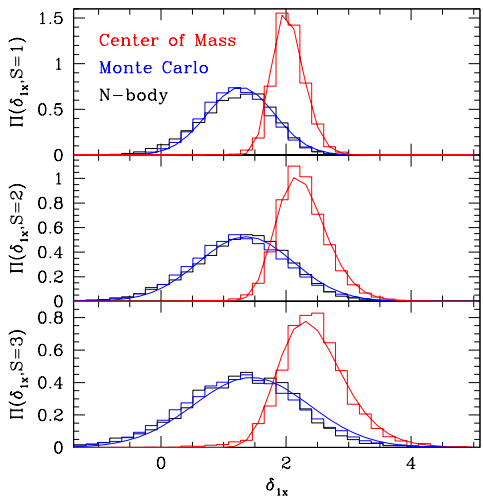}
\caption{Distribution of initial overdensities around randomly chosen halo particles when $S=1,2$ and $3$ (black histograms). Blue histograms show the first-crossing overdensities obtained from the same Monte-Carlos which were used to produce Fig.~\ref{fig1}. Smooth blue curves show our Eq.~(\ref{pd1x}) with parameters calibrated to fit Fig.~\ref{fig1}; i.e., it has no free parameters. Red histograms, which are more sharply peaked, show the same measurement but around halo centers-of-mass. In this case, smooth curves show a Lognormal with the same mean and rms as the measurements.}\label{fig2}
\end{center}
\end{figure}

Binning in $S$ yields the black histograms in Fig.~\ref{fig2}.  Our values of $S=1, 2$ and $3$ correspond to masses $M/10^{12}M_\odot =52,7.6$ and $2.2$ (each bin containing 2265, 31235, 5241 halos respectively).  The blue symbols show $\Pi(\delta_{1x},S)$ obtained from the same excursion set Monte-Carlo walks whose first crossing distribution is shown in Fig.~\ref{fig1}. The agreement between the two sets of histograms suggests impressive self-consistency of the excursion set approach. The numbers in the top half of the Table give a more quantitative comparison. We can see that the mean and variance of $\delta_{1x}$ from the Monte-Carlo walks are remarkably consistent with those from the N-body simulations. Notice that Eq.~(\ref{fig2}) slight overpredicts the mean compared to the Monte-Carlo; this is because Eq.~(\ref{fig2}) has been derived assuming the same smoothing for $\delta$ and $B$.

\medskip 

Choosing a random particle from each halo, rather than the one at the center of mass is crucial. The (more sharply peaked) red histograms in Fig.~\ref{fig2} show the initial overdensity associated with center-of-mass particles in the same halos, i.e. $\delta_{cm}$. These values tend to be larger than those for random particles, and they are almost always greater than $\delta_{sc}$. This is consistent with the analysis in \cite{SMT2001}, who argued that $\delta_{cm}$ in a halo is larger than for any other particle in the halo. In addition, the shape is quite non-Gaussian. The smooth curve shows a Lognormal which has the same mean and variance as the measurements; it provides a good description, in agreement with \cite{RKTZ2009}.
The bottom half of the Table gives these measured means and variances. Although they are in good agreement with \cite{RKTZ2009}, they are larger than those reported by \cite{DTS2013} who report that the rms scales as $0.2\,\sqrt{S}$.  It may be that our estimates of $\delta_{cm}$ (like those of \cite{RKTZ2009}) are broadened as a consequence of having assumed the initial volume to be spherical, whereas it is usually not \cite{DTS2013}.  

The statistics of $\delta_{1x}$ and $\delta_{cm}$ are different, the latter is more closely related to the predictions of the ellipsoidal collapse model of \cite{SMT2001}. For example, setting $p=0$ in Eq.~(3) of \cite{SMT2001} yields 
\begin{equation}
 \label{deltaec}
 \frac{\delta_{ec}}{\delta_{sc}} = 1 + 0.47\, 
  \left[5\frac{e^2\delta_{ec}^2}{S}\frac{S}{\delta_{sc}^2} \right]^{0.615}
\end{equation}
for the critical overdensity associated with the ellipsoidal collapse model of \cite{BM1996} ($e$ is a measure of the ellipticity of the shear field).  This shows that, at fixed $S$, the distribution of $e\delta/\sigma$ determines the distribution of $\delta_{ec}$.  Setting $p=0$ in Eq.~(A3) of \cite{SMT2001} yields $1/\sqrt{5}$ for the most probable value of $e\delta/\sigma$.  Therefore, the expression above suggests $\bar{\delta}_{ec}(S)\approx \delta_{sc} + 0.41 S^{0.615}$, which is in good agreement with the corresponding $\bar{\delta}_{cm}(S)$ values (bottom line of the Table). Similarly, the variance around the mean of $e\delta/\sigma$ is 0.14 \cite{DTS2013} so the rms of $(e\delta/\sigma)^{1.23}\approx 0.17$, making the expected variance of $\delta_{ec}\sim 0.032\,S$. It is possible that the discrepancy between theory and measurements, which is still of order a factor of 2 or 3 in the variance, may be reduced if one consider initial ellipsoidal patches rather than spherical ones.

\begin{widetext}
\begin{footnotesize}
\begin{tabular}{|c|c|c|c|}
  \hline
       & S=1 & S=2 & S=3\\
  \hline
N-body & $\bar{\delta}_{1x}=1.19\pm0.008;\rm{var}=0.34\pm0.007 $ 
       & $\bar{\delta}_{1x}=1.28\pm0.004;\rm{var}=0.59\pm0.005 $ 
       & $\bar{\delta}_{1x}=1.25\pm0.020;\rm{var}=0.90\pm 0.02 $  \\
Monte Carlo & $\bar{\delta}_{1x}=1.19\pm0.005;\rm{var}=0.29\pm0.005 $ 
            & $\bar{\delta}_{1x}=1.24\pm0.007;\rm{var}=0.58\pm0.007 $ 
            & $\bar{\delta}_{1x}=1.27\pm0.010;\rm{var}=0.85\pm 0.01 $  \\
Theory & $\bar{\delta}_{1x}=1.27;\rm{var}=0.29 $ 
       & $\bar{\delta}_{1x}=1.36;\rm{var}=0.57 $ 
       & $\bar{\delta}_{1x}=1.44;\rm{var}=0.85 $ \\  
\hline
N-body & $\bar{\delta}_{cm}=2.05\pm0.005;\rm{var}=0.07\pm0.007 $ 
       & $\bar{\delta}_{cm}=2.26\pm0.003;\rm{var}=0.17\pm0.004 $ 
       & $\bar{\delta}_{cm}=2.44\pm0.020;\rm{var}=0.30\pm0.010 $ \\
Theory & $\bar{\delta}_{ec}=2.09;\rm{var}=0.03$ 
       & $\bar{\delta}_{ec}=2.31;\rm{var}=0.06$ 
       & $\bar{\delta}_{ec}=2.49;\rm{var}=0.09$\\
\hline
\end{tabular}
\end{footnotesize}\\
\end{widetext}

A key point of our analysis is the distinction between the distribution of initial overdensities of center-of-mass halo particles and random ones. This was overlooked in \cite{RKTZ2009}, who have mistakenly claimed the inconsistency of the excursion set ansatz.  

Here, we have shown that once the statistical assumptions underlying the excursion set approach are considered and despite the differences with respect to the ellipsoidal collapse model prediction, a simple drifting diffusive barrier can provide a remarkable self-consistent description of the halos mass function as well as the distribution protohalo patches in simulations. Therefore, we believe our results motivate further studies along these lines. 

For example, we are not yet able to predict the (approximately Lognormal) statistics of center of mass walks from those of all walks, but see \cite{CS2013} for a first step in this direction. Also, because we focus on the statistics of all walks, and the effective barrier for these, our approach is complementary to that pursued by \cite{MS2012,PS2012,PSD2012}, who focus instead on developing an excursion set model for center-of-mass walks. In addition, there is a small but systematic tendency for the black histograms in Fig.~\ref{fig2} to peak at larger $\delta_{1x}$ than the blue as $S$ decreases. This may be a consequence of our having forced the barrier to be linear in $S$, or of using different filters for $\delta$ and $B$, or an artifact of the numerical halo detection algorithm. We are exploring this further. 
Such issues highlight the question of whether the Excursion Set approach is merely an effective one or if it provides the stochastic master equation description of the gravitational N-body problem. For now, this remains an open question which we hope to address in the future.

\medskip

\begin{acknowledgments}
I. Achitouv acknowledges support from the Trans-Regional Collaborative Research Center TRR 33 ``The Dark Universe'' of the Deutsche Forschungsgemeinschaft (DFG) and J. Weller. Y. Rasera thanks I. Balmes 
for fruitful discussions. R.K. Sheth is supported in part by NSF 0908241 and NASA NNX11A125G, and thanks E. Castorina for discussions. P.S. Corasaniti is supported by ERC Grant Agreement no. 279954.

\end{acknowledgments}

\end{document}